\documentclass[12pt]{article}
\usepackage{amsmath,amsfonts,amsthm,amscd,graphics}
\usepackage{palatino}    
\usepackage[mathcal]{euler}   

\newtheorem{thm}{Theorem}
\newtheorem{lemma}[thm]{Lemma}

\theoremstyle{definition}
\newtheorem*{definition}{Definition}

\newcommand{\cs}{C${}^*$}
\DeclareMathOperator{\tr}{tr}

\DeclareMathOperator{\Aut}{Aut}

\newcommand{\onto}{\to\mkern-14mu\to}
\DeclareMathOperator{\ad}{ad}
\newcommand{\co}{\mathbb{C}}
\newcommand{\A}{\mathcal{A}}
\newcommand{\Hi}{\mathcal{H}}
\newcommand{\m}{\varphi}
\newcommand{\Li}{\mathcal{L}}
\newcommand{\isom}{\mathrel{\widetilde\longrightarrow}}
\newcommand{\into}{\hookrightarrow}
\newcommand{\C}{\mathcal{C}}

\begin{document}
\begin{flushright}
\vspace*{-0.5in}
\begin{tabular}{l}
\textsf{\small hep-th/0302111}\\
\textsf{\small SISSA 16/2003/MP}\\
\textsf{\small CGPG-03/5-5}\\
\end{tabular}
\vspace{0.25in}
\end{flushright}
\title{Evolution in Quantum Causal Histories}
\author{
Eli Hawkins\thanks{Email address: hawkins@sissa.it}\\
 Scuola Internazionale Superiore di Studi Avanzati,\\
 Via Beirut 4, I-34014, Trieste, Italy\\
\and
Fotini Markopoulou\thanks{Email address: fotini@perimeterinstitute.ca}\\
 Perimeter Institute for Theoretical Physics,\\
 35 King Street North, Waterloo, Ontario N2J 2W9, Canada, and \\
 Department of Physics, University of Waterloo,\\
 Waterloo, Ontario N2L 3G1, Canada\\
\and
Hanno Sahlmann\thanks{Email address: hanno@gravity.psu.edu}\\
 Center for Gravitational Physics and Geometry, \\
 The Pennsylvania State University, \\
 University Park, PA 16802-6300, USA}
\date{}
{
\renewcommand{\newpage}{}
\maketitle
}
\begin{abstract}
We provide a precise definition and analysis of quantum causal histories (QCH's). A QCH consists of a discrete, locally finite, causal pre-spacetime with matrix algebras encoding the quantum structure at each event. The evolution of quantum states and observables is described by completely positive maps between the algebras at causally related events. We show that this local description of evolution is sufficient and that unitary evolution can be recovered wherever it should actually be expected. This formalism may describe a quantum cosmology without an assumption of global hyperbolicity; it is thus more general than the Wheeler-De\,Witt approach.  The structure of a QCH is also closely related to quantum information theory and algebraic quantum field theory on a causal set.
\end{abstract}

\section{Introduction}

The quest for a quantum theory of gravity is now several decades old and has been studied with approaches ranging from string theory and non-commutative geometry to loop quantum gravity, spin foams, and discrete models of Planck-scale spacetime.   Most of these take different interpretations of what a quantum theory of gravity should achieve, but with one thing in common: that it should be some merger of general relativity and quantum theory.   The final goal has not yet been achieved, but all these approaches have brought to us new understandings of the problem.

We come to the motivation for this paper by putting together a consistent list of features, selectively chosen from among the various indications of what a quantum theory of gravity may be.  Such a list is by no means unique, but we will find that it is fruitful;  it leads to a new notion of quantum cosmology and a new formulation of evolution in a Planck-scale theory.

Our list is the following:
1) At energies close to the Planck scale, the description of spacetime as a continuous,  $3+1$-dimensional manifold breaks down; this is the usual idea  for regularization  of quantum field theory.
2) Causality is still present in Planck-scale geometry. 
3) Quantum theory is valid in Planck-scale physics.
 4) In a fundamental theory of the universe, there should be a finite number of degrees of freedom in any finite region of the universe;  this is supported by black hole entropy calculations from both string theory and loop quantum gravity, and Bekenstein's physical arguments.
5) The theory should be background independent.

Quantum causal histories (QCH) were introduced in \cite{Mar00} as a formalism for quantum cosmology that realizes most of the above assumptions.  The idea is to use a causal set (locally finite, partially ordered set) as the description of the causal structure of the universe. Quantum theory is present through the assignment of  finite-dimensional Hilbert spaces to these events. They can be regarded as representing the fundamental building blocks of the universe at the Planck scale.  The composite Hilbert space for two spacelike separated events is the tensor product of the individual Hilbert spaces.  Evolution is implemented by unitary operators between complete pairs, special cases of sets of spacelike separated events which can be understood as isolated Planck-scale systems, or the discrete equivalent of successive Cauchy surfaces of an isolated component of spacetime. 

Clearly, this implements the first four properties above. The fifth is more difficult (and will not be resolved here). Because it involves a fixed causal structure, a QCH is not a completely background independent description. Nevertheless, all other degrees of freedom are quantum, and with the algebraic treatment presented here this may be a step toward a truly background independent model.

Quantum causal histories can also describe causal spin foam models (see the review \cite{Per}).  These are pre-spacetime models; that is, gravity and the familiar spacetime manifold are meant to be derived as the low-energy continuum approximation of such models.  A particular example of a QCH is a causal spin network \cite{Mar97,MS} and this was a principal motivation for the introduction of QCH's. A variation on the idea of a QCH was applied to a spin foam model in \cite{LO}.

However, as was already discussed in \cite{Mar00}, the definition of QCH there does not adequately take into account the causal structure of the underlying causal set.  In particular, the unitary evolution operators in that definition are indifferent to the individual relations interpolating between the ingoing and outgoing events.  The same unitary evolution is admitted regardless of these individual relations. This means that the quantum evolution does not respect the underlying causal structure.  Further, given two related events, there might not be a complete pair containing them, in which case we cannot say anything about the relation between states at the two events.  These issues are discussed in detail in the next section.

In this paper we address all of the above issues by giving a revised definition for  a quantum causal history in which  the causal structure is faithfully reflected in the quantum causal history.  We find that doing so requires working with algebras of operators, instead of Hilbert spaces, on the events.  Following the above idea of local finiteness at the fundamental level, these algebras are simple matrix \cs-algebras.  We find that a causal relation in the causal set corresponds to a completely positive map in the QCH.   This is not surprising, since this is  the most general kind of physical transformation allowed by quantum theory.  We also show that these maps can be regarded as fundamental and the unitary operators of the original definition can be reconstructed from them with no loss of information.

As a quantum cosmology, a QCH is most naturally interpreted as describing the universe as a very large collection of open quantum mechanical building blocks.  Unitary evolution arises only where there is a complete pair, which is the discrete analogue of an isolated system.  This is much more general than a Wheeler-De\,Witt type of quantum cosmology.  That ``canonical'' approach to quantum gravity is strongly dependent on global hyperbolicity and Cauchy surfaces. In contrast, QCH's allow that hyperbolicity may dissolve in the quantum foam. This seems more likely to describe the apparent non-unitarity of black hole evaporation than does the more traditional approach.
A QCH is also especially suited for describing the evolution of localized quantities without having to refer to a path integral or Hamiltonian constraint for the entire universe.
Aside from cosmology, we see with hindsight that there are several further ways of interpreting a QCH;  we discuss these in Section \ref{QFT} and the conclusions.

Concretely, the present paper is an in-depth analysis of QCH's and  opens up several new possibilities for the concrete implementation of models of that type.  For example,  it is interesting, and potentially practical,  that evolution in a QCH is almost identical to the quantum information theoretic description of quantum noise.  Also, the continuous time analogue of the completely positive evolution in a QCH,
is a  master equation, already extensively used in quantum cosmology.  We comment on  these connections in the conclusions.

The outline of this paper is as follows.  In Section \ref{QCH},  we review the original definition of quantum causal histories given in \cite{Mar00}, as well as the definitions of the basic structures in a causal set that are needed for the rest of the paper.   One of our main results is that the relations in a causal set correspond to completely positive maps in the QCH.  In order to give the reader some intuition about these maps before they have to face the precise definition of a QCH and the following proof that unitary evolution can be reconstructed from them, we give in Section \ref{CPmaps} a basic introduction to completely positive maps and a simple example.  Section \ref{ACH} is the revised, complete definition of a QCH.  Although this was not a motivation for this work, this definition can be understood as an algebraic quantum field theory on a causal set, with the operator algebras being simple matrix \cs-algebras.  We explain this in Section \ref{QFT}.  In Section \ref{proof}, we show that nothing is lost in going from the unitary operators of the original definition to the completely positive maps, since the former can be reconstructed from the latter, up to irrelevant phase factors.   We discuss the interpretation of our results and future directions in the conclusions.

\section{Review of Quantum Causal Histories}
\label{QCH}

In this section we review quantum causal histories.  They were defined
in \cite{Mar00} as a general formalism for causal spin foams in
quantum gravity \cite{MS,Mar97,Gup}\footnote{We will not discuss
specific spin foam models in the present article.  For reviews and discussion of spin foams, see \cite{Bae,Ori,Mar02}.}.

An important and distinguishing feature of the geometry of spacetime (as opposed to space) is causal structure. If a spacetime is time-orientable and has no closed timelike curves, then its causal structure can be completely described as a partial order relation on the events (points) of the spacetime. The definition of the relation is that $x\leq y$ (``$x$ precedes $y$'') if there exists a future-directed non-spacelike curve from $x$ to $y$. This relation  is transitive, i.e., if $x\leq y$ and $y\leq z$ then $x\leq z$, because future-directed curves can be concatenated. The assumption that there exist no closed timelike curves means precisely that  $x\leq y$ and $y\leq x$ if and only if $x=y$. These are the defining properties of a \emph{partial order relation}.

We are concerned here with a discrete analogue of a smooth spacetime. A  \emph{causal set} $\C$ is a locally finite, partially ordered set. That is, for any two events $x,y\in\C$, there exist (at most) finitely many events $z\in\C$ such that $x\leq z\leq y$.

If either $x\leq y$ or $y\leq x$, then we say that the events $x$ and $y$ are \emph{related}. If two events are not related, then we say that they are \emph{spacelike separated} and write $x\sim y$.

An \emph{acausal set} is a subset $\xi\subset \C$ such that all events in $\xi$ are spacelike separated to one another, i.e., $x,y\in \xi \implies x\sim y$. Acausal sets play the role of spacelike hypersurfaces in causal sets.

\begin{figure}
\center{\includegraphics{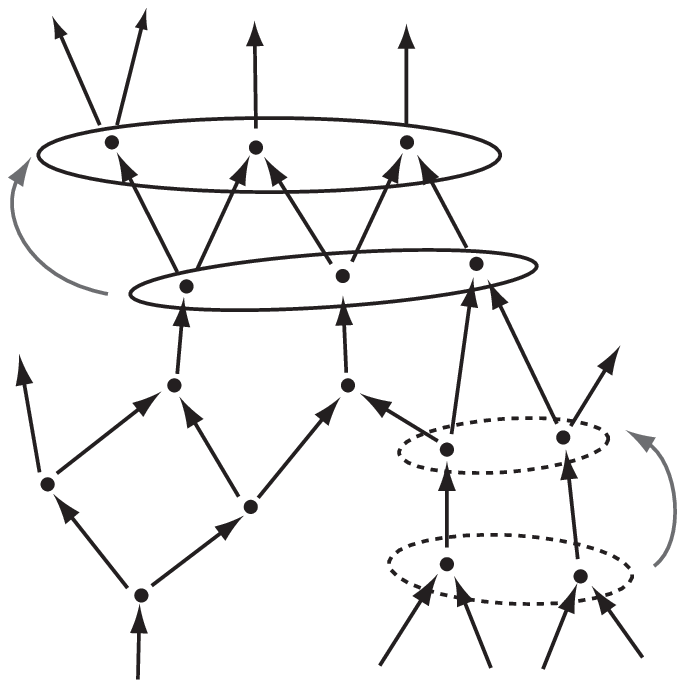}}
\caption{Two examples of complete pairs in a causal set.}
\label{cpairs}
\end{figure}

A causal set can usefully represented by the (directed) graph of elementary relations (Fig.~\ref{cpairs}). The set of vertices is simply $\C$. There is an edge $x\to y$ if $x\leq y$ but there does not exist any $z\in\C$ such that $x\leq z\leq y$. There is a simple analogue of a future-directed non-spacelike curve in a causal set. A \emph{future-directed path} is a directed path through the graph, that is, a sequence of events such that there exists an edge from each event to the next. A future-directed path is \emph{future (past) inextendible} if there exists no event in $\C$ which is in the future (past) of the entire path. Note that this graph generates the partial ordering: $x\leq y$ if and only if there exists a future directed path from $x$ to $y$.

From this, we can define complete future and past\footnote{These definitions are more restrictive than those given in previous papers.}.
An acausal set $\xi$ is a \emph{complete future} of an event $x$ if $\xi$ intersects any future-inextendible future-directed path that starts at $x$. Similarly, an acausal set $\xi$ is a \emph{complete past} of an event $x$ if $\xi$ intersects any past-inextendible future-directed path that ends at $x$.

In order to describe a local quantum structure on our causal set, we would like a (finite-dimensional) Hilbert space $\Hi(x)$ for every event $x\in\C$.
For two spacelike separated events $x\sim y\in\C$, the composite state
space for $\{x, y\}$ is $\Hi(\{x, y\})=\Hi(x)\otimes \Hi(y)$.  More generally,
for any acausal set $\xi\subseteq \C$,
$\Hi(\xi)=\bigotimes_{x\in\xi} \Hi(x)$.

In quantum mechanics, time evolution is unitary evolution between the Hilbert spaces at different times. When should we expect unitary maps between the Hilbert spaces for different acausal sets $\xi,\zeta\in\C$? Unitarity can be interpreted as meaning that no information is lost or added. We can only expect that no information is lost in evolving from $\xi$ to $\zeta$ if $\zeta$ is a complete future of $\xi$; otherwise, information coming from $\xi$ might ``miss'' $\zeta$. Likewise, we can only expect a state at $\zeta$ to be completely determined by one at $\xi$ if $\xi$ is a complete past of $\zeta$. When both of these conditions are satisfied --- $\xi$ is a complete past of $\zeta$ and $\zeta$ is a complete future of $\xi$ --- we call $\xi$ and $\zeta$ a complete pair and write $\xi\preceq\zeta$. We should expect a unitary map $\Hi(\xi)\to\Hi(\zeta)$ if and only if $\xi\preceq\zeta$. One can think of a complete pair as successive Cauchy surfaces of an isolated component of spacetime, or of all spacetime (see Fig.~\ref{cpairs}).

This leads to the definition of a quantum causal history given in \cite{Mar00}. It consists of a causal set $\C$, a finite-dimensional Hilbert space $\Hi(x)$ for every $x\in\C$, and a unitary map $u(\xi,\zeta) : \Hi(\xi)\to\Hi(\zeta)$ for any complete pair $\xi\preceq\zeta$, such that for $\xi\preceq\upsilon\preceq\zeta$,
\[
u(\upsilon,\zeta) u(\xi,\upsilon) = u(\xi,\zeta) .
\]

With this definition, the unitary maps on the complete pairs tell us very
little about the causal relations between the two acausal sets.  There
can be many configurations of the interpolating causal relations in a
complete pair, for example:
\begin{center}
\includegraphics{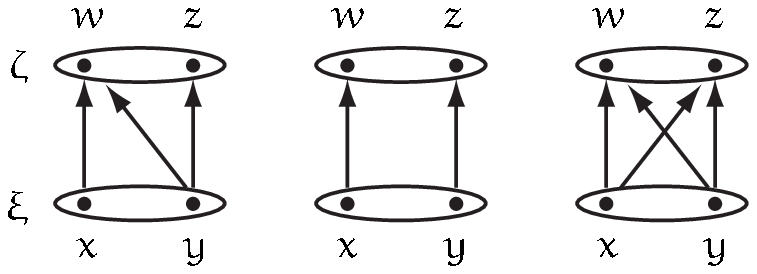}
\end{center}
The rules of a quantum causal history tell us that
there is a unitary map $u(\xi,\zeta):\Hi(\xi)\rightarrow \Hi(\zeta)$, but not how it should depend on the interpolating causal relations.

This is insufficient.
Consider for example the complete pair:
\begin{center}
\includegraphics{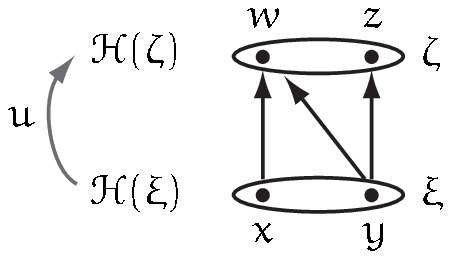}
\end{center}
If we start with an unentangled pure state $\lvert\Psi_x\otimes\Psi_y\rangle \in \Hi(\xi) = \Hi(x)\otimes\Hi(y)$, then this will evolve to
$\lvert\Psi'\rangle :=u(\xi,\zeta)\lvert\Psi_{x}\otimes\Psi_{y}\rangle\in \Hi(\zeta)$. The state at $z$ is described by a density matrix, $\rho_{z}=\tr_{w}\lvert\Psi'\rangle \langle\Psi'\rvert$.  With this causal structure, the state at $z$ should be completely determined by that at $y$. In other words $\rho_z$ should be independent of $\lvert\Psi_x\rangle$. However, with the above definition of a quantum causal history, there is no restriction on $u(\xi,\zeta)$ that would prevent $\rho_z$ from depending upon $\lvert\Psi_x\rangle$.

Another problem with this definition is the reliance on complete pairs. Given two related events $x\leq y$, there may not exist a complete pair containing these events. If so, with this definition of a quantum causal history, we do not know anything about the relation between states at $x$ and $y$. There should be some relation, and the concept of quantum causal history must be refined to provide this.

In this paper we address all of the above issues by giving a revised definition of  a quantum causal history in which  the causal structure of $\C$ is faithfully reflected in the quantum causal history.
We will define the QCH as an assignment of completely positive maps to pairs of causally related events. These maps relate   the local observables at the causally related events. In the next section we introduce the definition of a completely positive map on open quantum mechanical systems, in order to provide some intuition for the evolution in QCH before giving the new definition and results in Sections \ref{ACH} and \ref{proof}.

\section{Completely Positive Maps}
\label{CPmaps}

An ordinary quantum mechanical system may be isolated, or it may be coupled to some environment. The states of the system can be described with density matrices and the most general evolution that quantum mechanics allows for such a system is a linear, trace-preserving, completely positive (CP) map.  In this section we define CP maps and discuss their physical interpretation.

\begin{figure}
\center{
\includegraphics{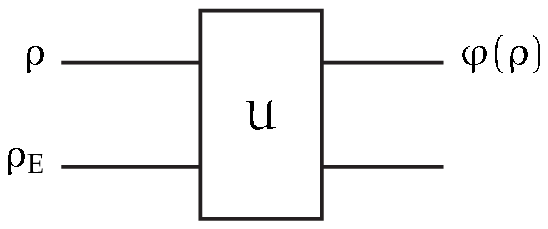}}
\caption{A system in a mixed state $\rho$ is coupled to some environment, $\rho_{E}$.  The joint system undergoes unitary evolution $U$.  From the output, $\m(\rho)$ is extracted, a system in a new mixed state which may or may not have the same dimensionality as the original system.  The transformation $\m$ is a linear, trace-preserving, completely positive map. }
\label{CPex}
\end{figure}

Consider a system $S$ in some mixed state given by $\rho$, a density matrix on its Hilbert space.  Let  $S$ interact with an environment $E$, also in a mixed state $\rho_E$.  The evolution of the joint system is given by a unitary operator $U$.  The process could be such that it is not feasible to extract the system $S$ again, but instead we extract $S'$, which has a density matrix $\rho'$.  See Fig.\ref{CPex}.  The transformation from $\rho$ to $\rho'$ is given by a map
\[
\m:\rho\rightarrow \rho'
\]
of the form
\[
\m(\rho)=\tr_{\mathrm{env}}\left[U(\rho\otimes\rho_{E})U^*\right].
\]

Such generic transformations from a quantum system to another are maps from the set of density matrices on $S$ to the set of density matrices on $S'$ that satisfy the following conditions:

\emph{Linear on the set of density matrices:} a probabilistic mixture of inputs leads to a probabilistic mixture of outputs, i.e., for probabilities $\{p_i\}$,
\[
\m\left(\sum_i p_i\rho_i\right)=\sum_i p_i\m(\rho_i).
\]

\emph{Trace-preserving:}  $\tr(\m(\rho))=\tr(\rho)=1$.  This simply means that  the normalization is preserved for any density matrix. The total probability for \emph{anything} to happen is, and remains, $1$.

\emph{Positive:} Probabilities are never negative, and so a density matrix has to be a positive operator (i.e., self-adjoint with only non-negative eigenvalues) in order to describe a physical system.  This is preserved by $\m$:  if $\rho$ is positive, then $\m(\rho)$ is also positive.  Note that this implies that $\m(a^*)=\m(a)^*$ for any matrix $a$.

\emph{Completely positive:}  It turns out that positivity is not enough for $\m$ to describe a physical process.  Suppose that we couple the system $S$ to another system $E$ that has trivial dynamics, $1_E$.  For all such trivial extensions, $1_E\otimes\m$ should also be positive.   This is the condition of complete positivity.
Complete positivity ensures that the evolution of $S$ is extendible in a trivial way to a physical evolution in a larger system $E\cup S$.  It is useful because it covers the case where $S$ is initially entangled with some distant isolated system\footnote{A standard example of an operation that is positive but not completely positive, and also not a physical operation on a system is taking the transpose of the density matrix.  See \cite{NC}, p.369.}.

A completely positive map on density matrices is equivalent to a completely positive map of observable operators in the opposite direction (in this case, backwards in time). A trace-preserving CP map of density matrices is dual to a unital\footnote{Unital means that $1\mapsto 1$.} CP map of observables.  The most general setting for completely positive maps is between \cs-algebras (see \cite{Lan}). In that context, there are many important tools for working with CP maps. In particular, the Stinespring construction shows that any unital CP map from a \cs-algebra to the algebra of bounded operators on a Hilbert space can be written in a universal form, as a representation of the \cs-algebra on a larger Hilbert space, composed with the projection onto the smaller Hilbert space.


\section{From Unitary to Completely Positive Maps}
 A quantum causal history can be reexpressed in terms of matrix algebras. Nothing is lost but trivial phase factors.

To any event $x\in \C$, we associate the algebra,
\[
\A(x):=\Li[\Hi(x)]
\mbox,\]
of matrices acting on $\Hi(x)$. Likewise, for any acausal $\xi\subset
\C$ there is the algebra,
\[
\A(\xi) := \Li[\Hi(\xi)] = \bigotimes_{x\in\xi} \A(x)
\mbox.\]
The inner products on the Hilbert spaces lead to involutions on the algebras. In this way, they can be regarded as \cs-algebras.

What do the unitary maps $u$ give us? For a complete pair
$\xi\preceq\zeta$, $u$ determines a $*$-isomorphism,
\[
\m(\xi,\zeta): \A(\zeta)\isom\A(\xi),\quad a\mapsto u^*(\xi,\zeta)\, a\,
u(\xi,\zeta)
\mbox.\]
If we are only interested in states (and not phase-factors), then $\m$
does not lose any important information from $u$.

Note that the maps $\m$ are directed backward in time. This is because we are regarding the matrix algebras as consisting of observable operators acting on the Hilbert spaces of local states. This is equivalent to a map of states (density matrices) in the other direction. Indeed, states should evolve into the future. However, because the unitary maps we are starting with are invertible, everything that we are doing can be expressed equivalently with time reversed.

\begin{figure}
\center{
\includegraphics{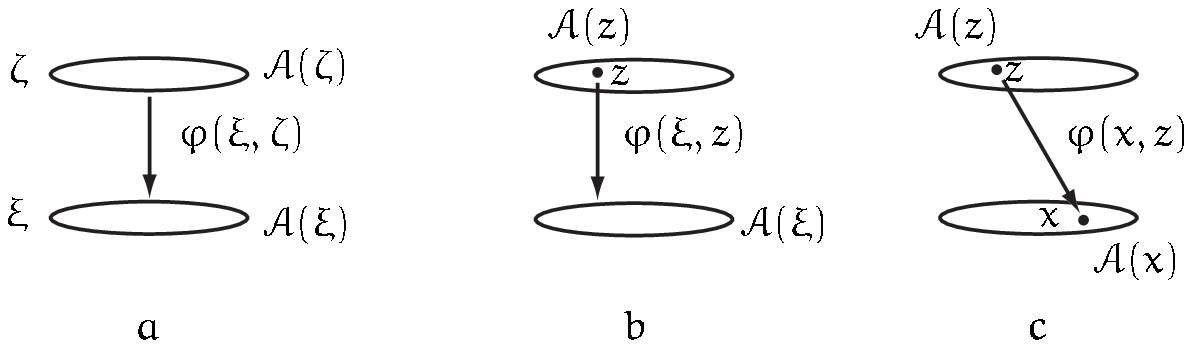}}
\caption{a) Unitary evolution for the complete pair $\xi\preceq \zeta$ gives a $*$-isomorphism $\m(\xi,\zeta)$ between the corresponding algebras. b) The acausal set $\xi$ is a complete past of $z$.  $\m(\xi,z)$ is a $*$-homomorphism.  c) In this case, $\m(x,z)$ is only a completely positive map.}
\label{maps1}
\end{figure}

What does $u$ give us more locally? If $z\in\zeta$, then $\A(z)$ can
be naturally thought of as a unital subalgebra of $\A(\zeta)$. Thus,
when $\xi\preceq\zeta$ and $z\in\zeta$, there is a unital
$*$-homomorphism, $\m(\xi,z):\A(z)\to\A(\xi)$, given by restricting
$\m(\xi,\zeta)$ to $\A(z)$. However, this is as far as homomorphisms
will go. For $x\in\xi$ and $z\in\zeta$ the naturally induced map
$\A(z)\to\A(x)$ is not a homomorphism; it is only completely positive.
See fig.\ref{maps1}.

Before describing the construction, we should note some properties of
the algebras that we are working with.  They are simple, finite-dimensional matrix algebras defined on Hilbert spaces. As such, they can be viewed in several ways and have convenient structures that are not available with most infinite-dimensional algebras. We have already noted that they are \cs-algebras. They are also von\,Neumann algebras, and because they are simple, they are (type I) von\,Neumann factors. We will make use of the trace on such an algebra; however, we will take advantage of the finiteness and use the normalized trace, given by the
formula,
\[
\tau(a):=\tr a/\tr 1 .
\]
The advantage of this will be to maintain the time-reversal symmetry of the formalism as explicitly as possible; the disadvantage is a slightly unorthodox normalization for density matrices ($\rho=1$ is correctly normalized).

Using the trace and the involution, we can construct an inner product on a matrix algebra,
\[
\langle a\vert b\rangle := \tau(a^*b)
\mbox,\]
Making the algebra into a Hilbert space as well. Since our algebras are Hilbert spaces, the linear maps among them have
adjoints.  In particular, if $v:A\to B$ is a linear map respecting the involution
$*$, the definition of its adjoint, $v^\dagger$ reduces to the
property that for any $a\in A$ and $b\in B$,
$\tau_A[a\,v^\dagger(b)]=\tau_B[v(a)b]$. We denote the adjoint of a map with a dagger to avoid confusion with the \cs-algebra involution $a\mapsto a^*$.

The adjoint of a homomorphism will not generally be a homomorphism,
however, it will be a completely positive (CP) map. The adjoint of any CP map is CP. If $v:A\to B$ is CP, and $A'\subseteq A$ and $B'\subseteq B$ are subalgebras, then define the \emph{reduction} of $v$ to $A'\to B'$ to be the CP map $i^\dagger_B\circ v\circ i_A$, where $i_A:A'\into A$ and $i_B:B'\into B$ are
the inclusions.
\[
\begin{CD}
A' @>{i_A}>> A \\
@. @VV{v}V \\
B' @>{i_B}>> B \\
\end{CD}
\]

\section{Quantum Causal Histories Redefined}
\label{ACH}

A quantum causal history thus leads to CP maps for pairs of
events. Specifically, if $\xi\preceq\zeta$, $x\in\xi$, and $z\in\zeta$,
then $\m(x,z):\A(z)\to\A(x)$ is the reduction of $\m(\xi,\zeta)$. Maps for pairs of related events give a more local description of a quantum causal history than maps for complete pairs of acausal sets. In a local description, we should regard
the maps $\m(x,z)$ as fundamental, and specify consistency conditions
on them. This leads to two questions, do local consistency conditions
exist, and is any information lost in going from global homomorphisms
to local CP maps? The answers, we will show, are yes and no.

\begin{figure}
\center{
\includegraphics{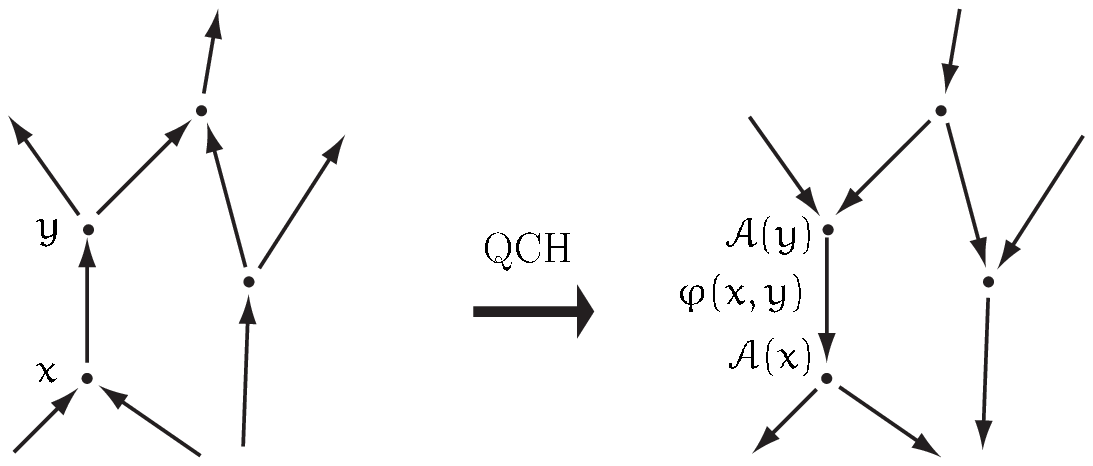}}
\caption{A quantum causal history associates algebras to events and completely positive maps to causal relations.}
\label{qchdef}
\end{figure}

There is one circumstance in which the adjoint of a homomorphism is
another homomorphism.
\begin{thm} 
   \label{unitary}
    If $v:A \isom B$ is an isomorphism, then
    $v^\dagger:B\to A$ is its inverse (also an isomorphism).
\end{thm}
\begin{proof}
    Homomorphisms in this category are trace-preserving. So, for $a\in
    A$ and $b\in B$,
    \[
    \tau_B[v(a)b] = \tau_B\left[v(a\,v^{-1}(b))\right]
    = \tau_A[a\,v^{-1}(b)]
    \mbox.\]
    Thus $v^{-1}=v^\dagger$.
\end{proof}
So, for $\xi\preceq\zeta$, $\m(\xi,\zeta)$ and $\m^\dagger(\xi,\zeta)$
are both isomorphisms. There is thus a time-reversal covariance; this
should be reflected in our axioms.

States on matrix algebras are all given by density matrices. Specifically, a density matrix is a positive matrix $\rho$ with $\tau(\rho)=1$; the expectation value of a matrix $a$ is $\tau(a\rho)$. The map of algebras $\m(x,y):\A(y)\to\A(x)$ induces a pullback map from states at $x$ to states at $y$. This is simply the adjoint $\m^\dagger(x,y)$ applied to density matrices. When the map of algebras is a
homomorphism, the map of states will take pure states to pure states.

Suppose that $\xi$ and $\zeta$ are acausal sets, not necessarily a
complete pair, but that $\zeta$ is a complete future of $\xi$. In
this case, no information should be lost in going from $\xi$ to
$\zeta$, and pure states should map to pure states. Thus, there
should be a homomorphism $\m(\xi,\zeta):\A(\zeta)\to\A(\xi)$.
Conversely, if instead $\xi$ is a complete past of $\zeta$, then
there should exist $\m(\xi,\zeta):\A(\zeta)\to\A(\xi)$ such that
$\m^\dagger(\xi,\zeta)$ is a homomorphism.

Suppose that $y\sim z$ and that $\xi$ is a complete past of both. The
homomorphisms $\m(\xi,y)$ and $\m(\xi,z)$ are restrictions of
$\m(\xi,\{y,z\})$. So, for $a\in\A(y)$ and $b\in\A(z)$,
\[
\m(\xi,y)(a)\cdot\m(\xi,z)(b) = \m(\xi,\{y,z\})(a\otimes b) =
\m(\xi,z)(b) \cdot \m(\xi,y)(a)
\mbox.\]
Thus the images of $\m(\xi,y)$ and $\m(\xi,z)$ must commute.

If such $\m$'s exist, and the $\m(x,y)$'s are reductions of them,
then they will satisfy the following properties.

\begin{definition}
    A quantum causal history consists of a simple matrix \cs-algebra $\A(x)$ for every event $x\in\C$ and a completely positive map $\m(x,y) : \A(y)\to\A(x)$ for every pair of related events $x\leq y$, satisfying the following axioms.
\end{definition}

\begin{figure}
\center{
\includegraphics{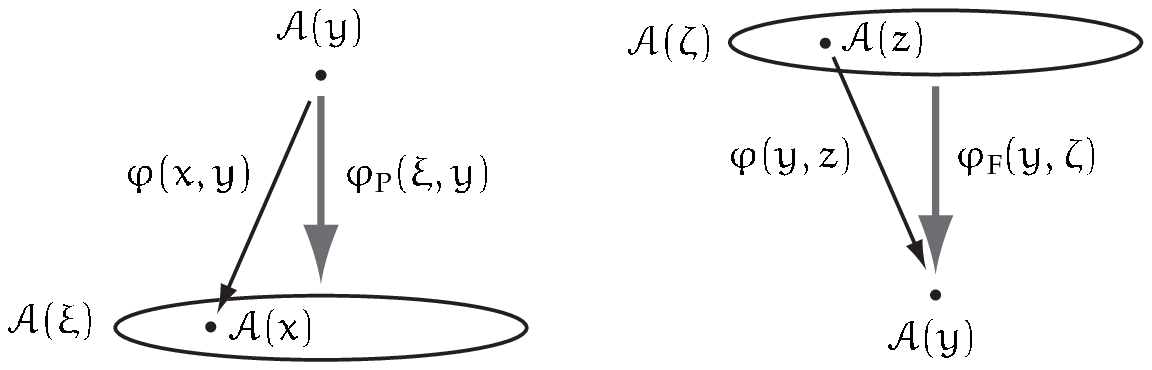}}
\caption{Extension.  On the left-hand side, the reduction of $\m_{\mathrm P}(\xi,y)$ to $\A(x)$ is $\m(x,y)$. On the right-hand side, the reduction of $\m_{\mathrm F}(y,\zeta)$ to $\A(\zeta)$ is $\m(y,z)$. }
\label{ax1}
\end{figure}

\emph{Axiom 1:} (Extension) For any $y\in \C$ and $\xi\subset
\C$ a complete past of $y$, there exists a homomorphism
\[
\m_{\mathrm P}(\xi,y) : \A(y)\to\A(\xi)
\mbox,\]
such that for each $x\in\xi$, the reduction of $\m_{\mathrm P}(\xi,y)$ to
$\A(x)\subset\A(\xi)$ is $\m(x,y)$.
Likewise, for any $\zeta\subset \C$ a complete future of $y$, there
exists a map
\[
\m_{\mathrm F}(y,\zeta) : \A(\zeta)\to\A(y)
\mbox,\]
such that $\m^\dagger_{\mathrm F}(y,\zeta)$ is a homomorphism and for
any $z\in\zeta$, the reduction (restriction) of $\m_{\mathrm
F}(y,\zeta)$ to $\A(z)\to\A(y)$ is $\m(y,z)$. See Fig.\ref{ax1}

\begin{figure}
\center{\includegraphics{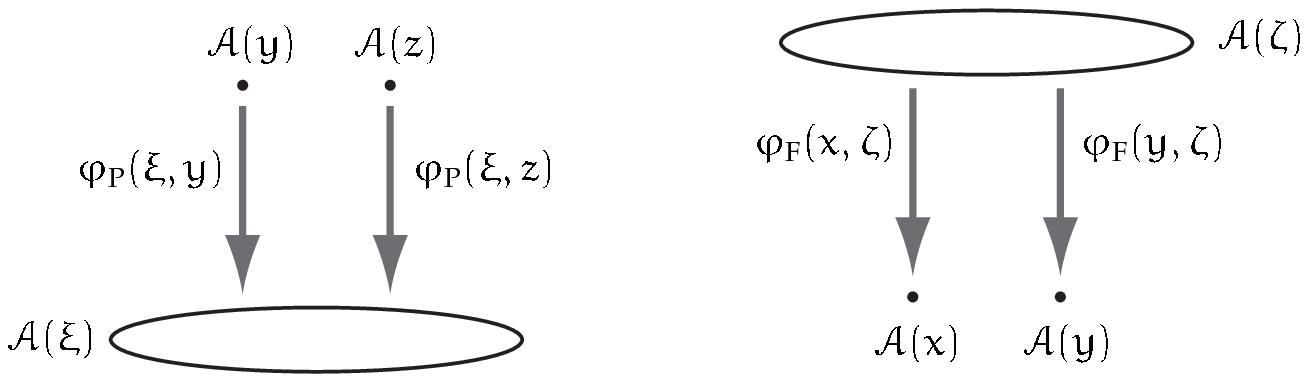}}
\caption{Spacelike commutativity. One the left-hand side, the images of  $\m_{\mathrm P}(\xi,z)$ and $\m_{\mathrm P}(\xi,y)$ (in $\A(\xi)$) commute.  On the right, the images of $\m_{\mathrm F}^\dagger(x,\zeta)$ and
$\m_{\mathrm F}^\dagger(y,\zeta)$ commute.}
\label{ax2}
\end{figure}

\emph{Axiom 2:} (Spacelike Commutativity) If $y\sim z\in \C$ and
$\xi\subset \C$ is a complete past of $y$ and $z$, then the images of
$\m_{\mathrm P}(\xi,z)$ and $\m_{\mathrm P}(\xi,y)$ (in $\A(\xi)$)
commute.  Likewise, if $\zeta\subset \C$ is a complete future of
$\{x,y\}$, then the images of $\m_{\mathrm F}^\dagger(x,\zeta)$ and
$\m_{\mathrm F}^\dagger(y,\zeta)$ commute. See Fig.\ref{ax2}.

\emph{Axiom 3:} (Composition) If $\zeta$ is a complete future of $x$ and a complete past of $y$, then $\m(x,y) = \m_{\mathrm F}(x,\zeta)\circ\m_{\mathrm P}(\zeta,y)$.

\section{Quantum Fields on Causal Sets}
\label{QFT}
This notion of quantum causal history can also be motivated from the ideas of algebraic quantum field theory. Algebraic quantum field theory (see \cite{Buc,Haa}) is a general approach to quantum field theory based on algebras of local observables, the relations among them, and their representations. The correct choice of axioms is still a matter of research, but for our
purposes, these subtleties are largely irrelevant.

An algebraic quantum field theory associates a von\,Neumann algebra to each causally complete region of spacetime. This generalizes easily to causal sets. The following definitions are exactly the same as for continuous spacetime. For any subset $X\subset \C$, define the \emph{causal complement} as
\[
X' := \{y\in\C \mid \forall x\in X : x\sim y\}
\]
the set of events which are spacelike to all of $X$. The \emph{causal completion} of $X$ is $X''$, and $X$ is \emph{causally complete} if $X=X''$. A causal complement is always causally complete (i.e., $X'''=X'$).

In the most restrictive axiomatic formulation of algebraic quantum field theory there is a von\,Neumann algebra $\A(X)$ for every causally complete region; These all share a common Hilbert space. Whenever $X\subseteq Y$, $\A(X)\subseteq\A(Y)$. For any causally complete region $X$, $\A(X')$ is $\A(X)'$, the commutant of $\A(X)$. The algebra associated to the causal completion of $X\cup Y$ is generated by $\A(X)$ and $\A(Y)$.

Some of the standard arguments about the properties of the local von\,Neumann algebras are valid for causal sets; some are not. The algebras should all be simple (i.e., von\,Neumann factors) because the theory would otherwise have local superselection sectors. For continuous spacetime it is believed that the local algebras should be type III${}_1$ hyperfinite factors; however, the reasoning involves the assumption that there exists a good ultraviolet scaling limit. This does not apply here; the small-scale structure of a causal set is discrete and not self similar at all.

Instead, it seems reasonable (following property 4 in the introduction) that only a finite amount of structure should be entrusted to each event. In other words, each von\,Neumann algebra should be a finite-dimensional matrix algebra. In von\,Neumann algebra terms, these are finite type I factors.

Consider the causal completion $\xi''$ of a finite acausal set $\xi$;  $\A(\xi'')$ should be generated by the algebras $\A(x)$ for $x\in\xi$. In fact, these algebras should commute, and therefore
\[
\A(\xi'') =\A(\xi) = \bigotimes_{x\in\xi}\A(x)
.
\]
Evolution by homomorphisms is sometimes possible. If $\xi$ and $\zeta$ are acausal sets such that $\zeta\subset \xi''$, then $\zeta''\subseteq\xi''$ and so $\A(\zeta)\subseteq\A(\xi)$. This inclusion is the evolution homomorphism when $\xi$ is a complete past of $\zeta$.

Not surprisingly, simple matrix algebras are much easier to work with than type III von\,Neumann factors. Using the (unique) normalized trace, any state is given by a density matrix. Recall that the adjoint maps $\m^\dagger(x,y)$ in a quantum causal history are the induced maps on density matrices.

Suppose that $\rho\in\A(x)$ is a density matrix at $x$. The density matrix $\m^\dagger(x,y)(\rho)\in\A(y)$ is the best approximation to $\rho$ among density matrices at $y$, in the sense that it minimizes the trace norm of the difference. The trace norm metric on density matrices is equal to the metric on states induced by the operator norm on observables.

So, we see that the obvious notion of an algebraic quantum field theory on a causal set, with the physically reasonable assumption of finite algebras on events, gives the structure of a QCH. The two concepts intersect, but are not equivalent. 

One could ignore the arguments about entropy \emph{et cetera} and consider a field theory with infinite algebras on events; however, depending upon the choice of algebras, even this might be described by some suitable generalization of a QCH. Likewise, a given QCH cannot necessarily be derived from an algebraic quantum field theory.

This does mean that the structure of a QCH encompasses a reasonable notion of a quantum field theory, and hence is capable of describing matter degrees of freedom. It also indicates how quantum fields on curved spacetime might be obtained as a limit of some quantum gravity model based on QCH's. Finally, this gives the skeptical reader another possible motivation for considering quantum causal histories.

\section{Unitary Maps from CP Maps}
\label{proof}

In this section we show that, if we start with the CP maps on the causal relations as the fundamental evolution operators, we can recover the unitary operators between the complete pairs.

The main result is that the axioms in the definition of a quantum causal history imply the following theorem:
\begin{thm}\label{main}
    For any acausal sets  $\xi,\zeta\subset \C$, if $\zeta$ is a complete
    future of $\xi$ or $\xi$ is a complete past of $\zeta$ then there
    exists a unique map
    \[
    \m(\xi,\zeta):\A(\zeta)\to\A(\xi)
    \]
    such that
    \begin{enumerate}
	\item For any $x\in\xi$ and $z\in\zeta$, the reduction of
	$\m(\xi,\zeta)$ to $\A(z)\to\A(x)$ is $\m(x,z)$.
	\item If $\xi$ is a complete past of $\zeta$, then $\m(\xi,\zeta)$ is
	a homomorphism.
	\item If $\zeta$ is a complete future of $\xi$, then
	$\m^\dagger(\xi,\zeta)$ is a homomorphism.
	\item If $\xi\preceq\zeta$ is a complete pair, then $\m(\xi,\zeta)$ is
	an isomorphism.
\item If $\xi\preceq\upsilon\preceq\zeta$, then $\m(\xi,\upsilon)\circ \m(\upsilon,\zeta)=\m(\xi,\zeta)$.
    \end{enumerate}
\end{thm}

In order to prove this, we first need to prove the uniqueness aspect.
Let $A, B_1, B_2$, \emph{et cetera} be (finitely many) simple,
finite-dimensional, matrix algebras. Let $B=\bigotimes_k B_k$. Any
unital homomorphism $A\to B$ has a reduction $A\to B_k$ for each $k$. We need
to show that two distinct homomorphisms will have distinct
reductions.

Let $\iota:A\to B$ be some unital homomorphism. Any other unital homomorphism can
be obtained from this by composing it with an automorphism of $B$.
All homomorphisms in this category are injective, so to simplify
notation, regard $A$ as a subalgebra of $B$ and $\iota$ as the
natural inclusion. First we prove the result at the Lie algebra level.

Denote the canonical inclusions by $j_k:B_k\to B$.  The reductions
of $\iota$ are simply $j_k^\dagger\circ\iota$. For any $\beta\in B$, define
$\ad_\beta:B\to B$ by $\ad_\beta(a)=[\beta,a]_-$.
\begin{lemma}\label{infinitessimal}
For any $\beta\in B$, if $j_k^\dagger\circ\ad_\beta\circ\iota=0$ for all
$k$, then $\ad_\beta\circ\iota=0$.
\end{lemma}
\begin{proof}
Let $C\subset B$ be the subspace spanned by the
$B_k$'s.  By the hypothesis, for any $a\in A$, $[\beta,a]_-$ must be
orthogonal\footnote{We are using the structure of $B$ as a Hilbert space.} to every $B_k$, and thus to $C$. Hence for any $c\in C$,
\[
0 = \tau(c[\beta,a]_-) = \tau (\beta [a,c]_-)
\mbox.\]
So $\beta\in [A,C]^\perp$. We need to
show that $[A,C]^\perp\subseteq A'$ (the commutant of $A$). The
orthogonal complement of $A'$ is $[A,A]$, so we simply need to prove
that $[A,A]\subseteq [A,C]$.

$C$ is closed as a Lie algebra and acts unitarily on $B$ by $\ad$.
$C$ generates $B$ as an algebra, therefore the only part of $B$
commuting with $C$ is the center, $\co\subset B$.  This is thus the
only trivial $C$-subrepresentation in $B$, and so $[B,C]=[B,B]$.
$[A,C]=\ad_C(A)$ is the sum of the nontrivial irreducible
$C$-representations which intersect with $A$.  Thus indeed
$[A,A]\subseteq [A,C]$, and the claim is proven.
\end{proof}

\begin{lemma}\label{reconstruction}
    Any homomorphism distinct from $\iota$ will have reductions
    distinct from those of $\iota$.
\end{lemma}
\begin{proof}
Let $H:=\{h\in \Aut B \mid j_k^\dagger\circ h\circ\iota=
j_k^\dagger\circ\iota\; \forall k\}$ be the subgroup that leaves the
reductions of $\iota$ invariant and $B^H$ the subalgebra of
$H$-invariant elements.  $\Aut A'\subset\Aut B$ leaves $\iota$
invariant, so $\Aut A'\subseteq H$ and $B^H\subseteq B^{\Aut A'} = A$.
$H$ is compact, so we can construct the orthogonal projection $p:B\onto
B^H\subset B$ by averaging over $H$.  Each map
$j_k^\dagger\circ\iota$ factors through $p$,
\[
j_k^\dagger\circ\iota = j_k^\dagger\circ p\circ\iota
\mbox.\]

Let $\beta\in A$ be any element commuting with $B^H$.
 For any $a\in A$ and $b\in B^H$,
\[
\tau(b[\beta,a]_-)=\tau(a[b,\beta]_-)=0
\mbox,\]
so $p\circ\ad_\beta=0$.
Thus
\[
j_k^\dagger\circ\ad_\beta\circ\iota = j_k^\dagger\circ\iota\circ\ad_\beta = j_k^\dagger\circ
p\circ\iota\circ\ad_\beta = j_k^\dagger\circ p\circ\ad_\beta\circ \iota =0
\mbox.\]
Therefore $\beta\in A\cap A' =\co$.
This implies that $B^H=A$.

Any other homomorphism can be written as $\alpha\circ\iota$
with $\alpha\in\Aut B$.  If this is distinct from $\iota$ then
$\alpha\not\in\Aut A$, but if $\alpha\circ\iota$ has the same
reductions as $\iota$ then $\alpha\in H$.  We would then have $H$
larger than just $\Aut A'$ and thus $B^H$ would have to be a \emph{proper}
subalgebra of $A$.
\end{proof}

\begin{proof}[Proof of Theorem~\ref{main}]
Lemma \ref{reconstruction} implies the uniqueness of the $\m_{\mathrm
P}(\xi,y)$ in Axiom 1. If $\xi$ is a complete past of $\zeta$, then if
a homomorphism $\m_{\mathrm P}(\xi,\zeta):\A(\zeta)\to\A(\xi)$ exists
with the $\m_{\mathrm P}(\xi,z)$'s (for $z\in\zeta$) as its reductions, it can be
reconstructed by multiplying the $\m_{\mathrm P}(\xi,z)$'s together,
so $\m_{\mathrm P}(\xi,\zeta)$ is unique. Conversely,
because of Axiom 2, we \emph{can} multiply together the $\m_{\mathrm
P}(\xi,z)$'s to define $\m_{\mathrm P}(\xi,\zeta)$.

Analogously, for $\zeta$ a complete future of $\xi$, a unique map
$\m_{\mathrm F}(\xi,\zeta)$ exists, satisfying the requirements of
Property 3.

Suppose that both of these are true.  That is, $\xi\preceq\zeta$ is a
complete pair.  We need to show that $\m_{\mathrm P}(\xi,\zeta) =
\m_{\mathrm F}(\xi,\zeta)$ and is an isomorphism.

Because all homomorphisms in this category are injective, the
existence of homomorphisms in both directions between $\A(\xi)$ and
$\A(\zeta)$ shows that they are of equal dimension and therefore
isomorphic. These homomorphisms must therefore be isomorphisms.

So, $\m^\dagger_{\mathrm F}(\xi,\zeta)$ is an isomorphism, and by
Theorem~\ref{unitary}, $\m_{\mathrm F}(\xi,\zeta)$ is its inverse.
The reductions of $\m_{\mathrm F}(\xi,\zeta)$ are the same as for
$\m_{\mathrm P}(\xi,\zeta)$ and as it is a homomorphism, the above
uniqueness implies $\m_{\mathrm P}(\xi,\zeta) = \m_{\mathrm
F}(\xi,\zeta)$.

Now suppose that $\xi\preceq\upsilon\preceq\zeta$. Let $x\in\xi$ and $y\in\zeta$. By what we have proven so far, $\m_{\mathrm F}(x,\upsilon)$ is a reduction of $\m(\xi,\upsilon)$. By Axiom 3, $\m(x,y) = \m_{\mathrm F}(x,\upsilon)\circ\m_{\mathrm P}(\upsilon,y)$. So, $\m(x,y)$ is a reduction of $\m(\xi,\upsilon)\circ\m_{\mathrm P}(\upsilon,y)$. By Lemma \ref{reconstruction}, since the reductions are the same, we must have that $\m_{\mathrm P}(\xi,y) = \m(\xi,\upsilon)\circ\m_{\mathrm P}(\upsilon,y)$. Using Lemma \ref{reconstruction} again, this shows that $\m(\xi,\upsilon)\circ \m(\upsilon,\zeta)=\m(\xi,\zeta)$.
\end{proof}
We have then shown that the system of isomorphisms associated to complete pairs can be reconstructed from the local information of CP maps associated to pairs of events. Because $\C$ is assumed to be locally finite, there is no problem in going from these isomorphisms of algebras to a system of unitary maps satisfying the original definition of a quantum causal history; the only arbitrariness is a choice of irrelevant phase factors.

On the other hand, the local formulation is restrictive. Not every choice of unitary maps for complete pairs can be expressed in terms of CP maps in this way. This formulation enforces local causality.

\section{Conclusions}

In this paper, we have provided a complete definition of a quantum causal history, a pre-spacetime framework for a quantum theory of gravity in which there is a simple matrix algebra of operators for every event in the universe (represented by a causal set) and a completely positive evolution map for every causally related pair of events.  These represent the transfer of quantum information between events in the quantum spacetime.  We showed that the axioms in the definition allow us to reconstruct the unitary evolution between complete pairs, a necessary step since complete pairs correspond to isolated systems.

In the picture of the universe as a QCH, it can be thought of as a very large collection of quantum-mechanical building blocks.  Everything is (locally) finite, both the dimensionality of these systems as well as the causal set.  We regard this as a simple way to model the idea that there is a finite number of fundamental degrees of freedom in a finite region of the universe.  Although there is certainly no hard evidence for this, such ideas are supported by Bekenstein's arguments, by the black hole entropy calculations from both string theory and loop quantum gravity, and is related to holographic ideas and the arguments for a many-Hilbert space M-theory in \cite{BF}.  Even independent of such motivations, finiteness allows significant simplifications which we have been able to exploit in this paper.

It may be possible to generalize a QCH to use some infinite-dimensional algebras. Theorem \ref{main} may extend to type II${}_1$ von\,Neumann factors, which have finite traces. However, we have not investigated this possibility in detail because the physical motivations for using matrix algebras are much stronger.

It is interesting to note the multiple ways that quantum causal histories can be motivated, including:
\begin{enumerate}
\item
 The original definition considering them as the general formalism for causal spin networks \cite{Mar00}.
 \item
As an algebraic quantum field theory on a causal set (Section \ref{QFT}).
 \item
 As an alternative to the usual quantum cosmology of a wavefunction of the universe evolving according to the Wheeler-De\,Witt equation.    A QCH is a framework for quantum cosmology that is not restricted to globally hyperbolic spacetimes, as is the case with a WDW-type quantum cosmology. This is discussed in more detail in \cite{Mar02}.
\item
 The local evolution in a QCH is well-suited for addressing the important problem of identifying localized propagating  degrees of freedom in a spin foam, without having to use the entire partition function.
 \item
  Precisely the same objects that we used (matrix algebras and CP maps) are basic tools in quantum information theory.  Conceptually, this suggests that the universe modeled as a QCH is nothing but a background-independent quantum computer.   More practically, the analogue of the much sought-after propagating ``particle'' in a spin foam, in quantum information theoretic language, is a quantum channel. Tools such as the operator-sum representation of CP maps, and even ideas from the theory of error-correction in quantum computation,  might be used to identify such objects.
\end{enumerate}

There is an important clarification we should make.  If quantum causal
histories are motivated via 2 above, then this is not gravity, but a kind of
quantum field theory on a peculiar background.  This is related to the
issue of the \emph{given} causal set that a QCH is built on.  If the
procedure is simply ``given a causal set $\C$, a QCH is constructed via the
procedure of Section \ref{ACH},'' then this cannot have much to do with a
dynamical theory of space-time.  Unless, somehow, all the gravitation
degrees of freedom are in the algebras and $\C$ is some auxiliary
background, but we are not aware of such a scheme.
In the quantum causal histories motivated from spin foams in quantum
gravity (4 above), there is an further step.  The partition function for a
(causal) spin foam model includes a sum over all causal sets.  This is the
usual path integral formulation of quantum gravity, with  each term in the path
integral being a QCH.  It would be very interesting to understand if there is
a way for a QCH to be a dynamical quantum space-time without the summation
of a path integral.  For this, one would need the causal set and the
algebras to ``interact'' with each other.

The reason why there are so many ways to understand quantum causal histories is partly the general nature of a CP map.  It is, after all, the most general physical transformation allowed by quantum theory.
Completely positive maps already appear in several places in quantum cosmology and we close this paper by referring to a few examples.

In the context of continuous time evolution, master equations are the familiar tools used to describe quantum noise, for example in quantum optics.  These are differential equations, governing the non-unitary evolution of the density matrices of dissipative quantum systems (e.g., \cite{Lin,CL}).  When a master equation is integrated, the evolution of the density matrix for a finite time-step is given by a completely positive map.

Another interesting comparison is between our CP maps and decoherence.  The term decoherence is used in the literature both to describe quantum noise, and the more specific process of phase damping that is believed to play a role in the emergence of classicality (see, for example, \cite{Zur}).  Using decoherence to mean the second, we note that it is possible to incorporate processes of that type in a quantum causal history. In \cite{BIP1,BIP2}, an interesting scheme for describing decoherence was proposed that is closely related to quantum causal histories.

In this new formulation, with the completely positive maps as the fundamental evolution operators, we obtain a quantum cosmology in which the universe is simply a very large collection of finite quantum systems. The information gets shuffled around, according to the causal structure.  Local observers can lose information simply because, by causality, it is not available to them.  This is much like information loss due to the presence of a horizon.

Finally, we note that there are several tools that can be used for the practical implementation of CP maps in a specific model.  These include the operator-sum representation that is frequently used in quantum information theory  (see \cite{NC}, chapter 8 and \cite{Sch} for excellent reviews,  and, for example,  \cite{TCDGS,Cho} for the case of dimensionality change from input to output),
and the Stinespring construction in operator algebras \cite{Lan}.

\section*{Acknowledgments}
We are very grateful to the Jesse Phillips Foundation for providing  the funds that brought us together in Berlin.
FM would like to thank Barbara Terhal for very useful  pointers to the relevant literature in quantum  information theory,  Mike Mosca and David Poulin for discussions on this subject, Lee Smolin and Olaf Dreyer for their comments on this work, and the Albert Einstein Institute, Golm, where part of this work was carried out. Part of this work was supported by NSF grant PHY-0090091.


\end{document}